\documentclass[conference]{IEEEtran}
\IEEEoverridecommandlockouts
\usepackage{cite}
\usepackage{amsmath,amssymb,amsfonts}
\usepackage{graphicx}
\usepackage{textcomp}
\usepackage{xcolor}
\usepackage{booktabs,multirow,url}
\usepackage{newtxtext,newtxmath}
\usepackage{flushend}
\usepackage[hidelinks]{hyperref}
\newcommand{\nSone}{12}

\newcommand{\SoneEbaseone}{0}

\newcommand{\SoneEbasetwo}{3}
\newcommand{\SoneEstricttwo}{5}
\newcommand{\SoneEbasethree}{5}
\newcommand{\SoneEstrictthree}{3}
\newcommand{\SoneEbasefour}{4}

\newcommand{\nStwo}{26}

\newcommand{\StwoEbaseone}{2}

\newcommand{\StwoEbasetwo}{7}
\newcommand{\StwoEstricttwo}{11}
\newcommand{\StwoEbasethree}{14}
\newcommand{\StwoEstrictthree}{10}
\newcommand{\StwoEbasefour}{3}

\newcommand{\nSthree}{12}

\newcommand{\SthreeEbaseone}{11}

\newcommand{\SthreeEbasetwo}{0}
\newcommand{\SthreeEstricttwo}{1}
\newcommand{\SthreeEbasethree}{1}
\newcommand{\SthreeEstrictthree}{0}
\newcommand{\SthreeEbasefour}{0}

\newcommand{\nEfour}{7}

\newcommand{\nEthreePlus}{27}
\newcommand{\nEthreePlusStrict}{20}
\newcommand{\nLfour}{128}
\newcommand{\nLfourStrict}{122}
\newcommand{\nLfourrecovered}{19}
\newcommand{\nLfourpartial}{54}
\newcommand{\nLfourfailed}{45}
\newcommand{\nLfourunclear}{10}
\newcommand{\SoneanyLfour}{11}

\newcommand{\StwoanyLfour}{25}

\newcommand{\MoneLfour}{27}
\newcommand{\Monerec}{1}
\newcommand{\Monefailed}{13}
\newcommand{\Monepartial}{13}
\newcommand{\MonegeOneSOT}{37}
\newcommand{\MonegeOneSthree}{5}
\newcommand{\MoneLthreeSthree}{1}

\newcommand{\MtwoLfour}{20}

\newcommand{\MtwogeOneSOT}{38}
\newcommand{\MtwogeOneSthree}{11}
\newcommand{\MtwoLthreeSthree}{2}

\newcommand{\MthreeLfour}{15}

\newcommand{\MthreegeOneSOT}{34}
\newcommand{\MthreegeOneSthree}{8}
\newcommand{\MthreeLthreeSthree}{6}

\newcommand{\MfourLfour}{0}

\newcommand{\MfourgeOneSOT}{2}
\newcommand{\MfourgeOneSthree}{6}
\newcommand{\MfourLthreeSthree}{3}

\newcommand{\MfiveLfour}{7}

\newcommand{\MfivegeOneSOT}{38}
\newcommand{\MfivegeOneSthree}{10}
\newcommand{\MfiveLthreeSthree}{6}

\newcommand{\MsixLfour}{22}
\newcommand{\Msixrec}{7}
\newcommand{\Msixfailed}{9}

\newcommand{\MsixgeOneSOT}{37}
\newcommand{\MsixgeOneSthree}{5}
\newcommand{\MsixLthreeSthree}{2}

\newcommand{\MsevenLfour}{12}

\newcommand{\MsevengeOneSOT}{37}
\newcommand{\MsevengeOneSthree}{3}
\newcommand{\MsevenLthreeSthree}{0}

\newcommand{\MeightLfour}{25}
\newcommand{\Meightrec}{7}

\newcommand{\MeightgeOneSOT}{38}
\newcommand{\MeightgeOneSthree}{10}
\newcommand{\MeightLthreeSthree}{2}

\newcommand{\agrRawE}{21}
\newcommand{\agrQwkE}{0.68}
\newcommand{\agrRawM}{298}
\newcommand{\agrQwkM}{0.80}
\newcommand{\agrQwkMfive}{0.51}
\newcommand{\nDisagree}{131}
\newcommand{\nDisagreeGapTwo}{71}
\newcommand{\nDisagreeBhigher}{98}
\newcommand{\nDisagreeSystems}{45}
\newcommand{\nOverrides}{10}
\newcommand{\nVerifyRequired}{87}
\newcommand{\nVerifyAudit}{30}
\newcommand{\nOutcomeAdj}{35}

\newcommand{\nGuar}{56}
\newcommand{\nGuarFormal}{12}
\newcommand{\nCand}{414}
\newcommand{\nDev}{21}

\newcommand{\nLfourIssueOnly}{98}

\begin{document}
\bstctlcite{IEEEexample:BSTcontrol}

\title{Exercised, Rarely Recovered: What Public Reports Say About Recovery in 50 AI Agent Harnesses and Runtimes}

\author{\IEEEauthorblockN{Kaviru Hapuarachchi}
\IEEEauthorblockA{University of Colombo School of Computing\\ Colombo, Sri Lanka\\ 2022is031@stu.ucsc.cmb.ac.lk}}

\maketitle

\begin{abstract}
When an AI coding agent deletes the wrong files or a long agent run crashes, users fall back on recovery mechanisms: resume from a checkpoint, undo, ask for approval, read the log. Research runtimes promise stronger guarantees, such as saga-style compensation and transactional rollback. We ask what public evidence shows about whether either kind works in real use. Under a protocol frozen before data collection, we graded \nSone{} persistent agent runtimes, \nStwo{} coding-agent harnesses and \nSthree{} research runtimes on eight recovery mechanisms, from \emph{not found} to \emph{exercised in a real run}, with two blind coding passes, adjudication and re-verification against archived sources. Recovery is exercised almost everywhere in practice: \SoneanyLfour{} of \nSone{} runtimes and \StwoanyLfour{} of \nStwo{} harnesses have at least one real-use report. Yet of \nLfour{} such reports, only \nLfourrecovered{} describe the run fully recovering; \nLfourpartial{} are partial and \nLfourfailed{} failed. Checkpoint/resume recovered once in \MoneLfour{} reports, rollback once in \MthreeLfour{}. Research runtimes show the opposite profile: guarantees evaluated on benchmarks, none reported in operation, and compensation of external effects, their central guarantee, is offered by no harness at all.
\end{abstract}


\section{Introduction}
A Gemini CLI user asked the agent to scaffold a project, clicked \emph{allow always} on a stuck package install, and watched it run \texttt{rm -rf} over their home folders.\footnote{\url{https://github.com/google-gemini/gemini-cli/issues/2617}} An Aider user ran \texttt{/undo} on a bad edit and lost their own uncommitted work with it.\footnote{\url{https://github.com/Aider-AI/aider/issues/513}} A Kilo Code user found that \emph{revert} undid the main agent's edits but not its sub-agents'.\footnote{\url{https://github.com/Kilo-Org/kilocode/issues/11963}} In each case a recovery mechanism did exactly what it was built for, on real work, and the user still lost something.

\begin{figure}[!t]
\centering
\includegraphics[width=\columnwidth]{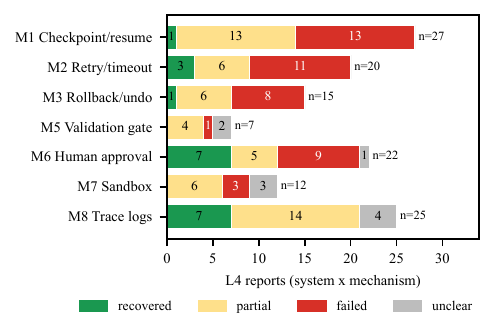}
\caption{Outcome of every public report of a recovery mechanism used in a real run (\nLfour{} reports, \nSone{}+\nStwo{} harnesses and runtimes). Bars are mechanism classes; M4 has no report. \textbf{Mechanisms that restore state (M1--M3) almost never fully recover the run; trace logs (M8) are the only class with no failures.}}
\label{fig:outcomes}
\end{figure}

Harnesses such as Claude Code, Codex and OpenHands now run long, side-effecting tasks for people who cannot read every step, so these mechanisms carry the weight of keeping agent work safe. Research has answered with stronger guarantees: saga-style compensation for multi-agent plans~\cite{chang2025sagallm,perera2026rac,geng2025alas}, transactional execution of agent actions~\cite{mohammadi2026atomix,chen2026cordon,sun2026agentictxn}, and checkpoint/restore of agent environments~\cite{wu2026crab,dong2026deltabox,yan2025ftsandbox}. But these are shown on benchmarks. Nobody has asked, across systems, whether the guarantees reach practice, or whether what harnesses actually ship holds up when it is used. Prior work covers parts of this: inventories of deployed agents~\cite{staufer2026agentindex}, maturity grades for agent repositories~\cite{naik2026agentstripes}, source-code anatomies of harnesses~\cite{barbaste2026harnessanatomy,liu2026diveclaudecode}, and one mechanism (persistence) in one framework~\cite{mastouri2026builttoremember}.

We answer with public evidence. We call a mechanism \emph{exercised} when an identifiable operator reports using it in a real run, and we record whether that run recovered. We ask:
\textbf{RQ1} How far does public evidence of production use go for harnesses, persistent runtimes and research runtimes?
\textbf{RQ2} For eight recovery mechanisms, how far does the evidence go, and when a mechanism is exercised, does it recover the run?
\textbf{RQ3} How do research guarantees compare with what harnesses ship and operators use?

\noindent Our contributions are:
\begin{itemize}
\item a recovery-specific \emph{mechanism $\times$ evidence-level} dataset for 50 systems, coded under a frozen protocol, with every label traced to an archived source;
\item the finding that exercised recovery is common but rarely complete: \nLfourfailed{} failed and \nLfourpartial{} partial of \nLfour{} real-use reports (Fig.~\ref{fig:outcomes});
\item a guarantee-to-practice map showing that compensation, research's strongest guarantee, has no counterpart in any harness, while no research runtime is shown in operation.
\end{itemize}

\section{Method}
We froze the protocol and codebook (SHA\nobreakdash-256 hashed and timestamped) before the first discovery query, and logged all \nDev{} departures in the replication package. We grade \emph{public evidence found within a bounded search}; a missing label means \emph{not found}, not \emph{not used}.

\subsection{Population}
Three strata were fixed in the protocol. \textbf{S1} contains the 12 open-source persistent agent runtimes of a predecessor study (e.g., OpenHands, Letta, Dify, n8n), carried over by rule and recoded fresh. \textbf{S2} contains coding-agent harnesses, open or proprietary, found as the union of three deterministic routes (GitHub search, package registries and curated lists); a harness had to operate the model loop itself and execute tools on the user's machine or workspace (criteria H1--H2). \textbf{S3} contains research agent runtimes that propose a recovery or reliability mechanism, found by predeclared scholarly queries plus seeds. Of \nCand{} candidates, \nStwo{} S2 and \nSthree{} S3 systems were included; each exclusion has one primary reason code. The evidence cutoff was 2026-11-09 23:59 (UTC+5:30).

\subsection{Coding scheme}
\emph{Production evidence (RQ1)} uses a six-step ladder reused unchanged from the predecessor study: E0 none, E1 benchmark or demonstration, E2 self-described use, E3 a named operator and workflow with corroboration, E4 an attributable sustained workflow with ownership, monitoring and recovery or incident evidence, and E5 E4 across multiple organisations with a measurement window. A vendor case alone is capped at E3 and self-description at E2.

\emph{Recovery mechanisms (RQ2)} are eight classes: M1 checkpoint/resume, M2 retry/timeout, M3 rollback/undo of the agent's own state, M4 compensation of effects already committed in external systems, M5 validation gates, M6 human approval/interruption, M7 sandboxing, and M8 trace logs usable for recovery. For each system and class we record the strongest evidence on an ordinal ladder: L0 not found, L1 documented, L2 demonstrated (example, test or tutorial), L3 evaluated (measured in a paper or benchmark), and L4 exercised (an operator-attributable report of the mechanism used in a real run or incident). Every L4 also records whether the run \emph{recovered}, \emph{partially} recovered, \emph{failed}, or is \emph{unclear}. Feature requests and statements of habit do not count as L4.

Each source is classed as project-, vendor-, operator-controlled or independent. We report a \emph{base} view and a \emph{strict} view, in which E3+ and L4 need at least one operator-controlled or independent source.

\subsection{Coding procedure}
Each system received a bounded search with a fixed list of query types and an archived snapshot of every cited source, with exact excerpts. Two passes coded all 50 systems independently: Pass~A and Pass~B were run by LLM-based coding agents on two different models (Pass~A: Claude Opus~5.5; Pass~B: OpenAI GPT-6 Astra), each blind to the other and to the predecessor study's labels, and each working only from the frozen codebook and its own archived sources. Agreement was \agrRawM{}/400 identical mechanism levels (quadratic-weighted $\kappa=\agrQwkM{}$; weakest M5, $\kappa=\agrQwkMfive{}$) and \agrRawE{}/50 identical E-levels ($\kappa=\agrQwkE{}$). The \nDisagree{} disagreements (\nDisagreeSystems{} systems; \nDisagreeGapTwo{} differing by two or more levels) were adjudicated blind: the adjudicator saw the coders as X and Y in random order and resolved each item against the codebook and the archived snapshots, not by majority. Pass~B was the higher of the two in \nDisagreeBhigher{} of the \nDisagree{} disputes.

The protocol then requires re-verification of every final L4, E3+ and every resolved disagreement of two or more levels. We built an evidence dossier for each of the \nVerifyRequired{} required items plus a fixed-seed audit sample of \nVerifyAudit{} agreed items, pre-screened them, and decided each item against the frozen rules; this changed \nOverrides{} labels (eight lowered, one raised, one L4 outcome changed). \nOutcomeAdj{} L4 items on which both passes agreed on the level but not on the outcome were resolved by the same blind procedure. We then hash-locked the dataset; every number in this paper is generated from it by script. The author designed the study, wrote and froze the protocol and codebook, defined every decision rule and the verification procedure, and directed the pipeline. LLM agents performed the two coding passes and subsequent adjudication and verification; after these stages, the author independently checked every final label and outcome against the cited evidence and codebook.

\section{Results}

\subsection{RQ1: production evidence}
\begin{table}[t]
\caption{Production-evidence level per stratum, base view (strict view in parentheses). \textbf{Harnesses and runtimes reach sustained use (E4); research runtimes stop at benchmarks (E1).}}
\label{tab:e}
\centering\small
\begin{tabular}{lrrrrr}
\toprule
Stratum & $n$ & E1 & E2 & E3 & E4\\
\midrule
S1 persistent runtimes & \nSone & \SoneEbaseone & \SoneEbasetwo{} (\SoneEstricttwo) & \SoneEbasethree{} (\SoneEstrictthree) & \SoneEbasefour\\
S2 coding harnesses & \nStwo & \StwoEbaseone & \StwoEbasetwo{} (\StwoEstricttwo) & \StwoEbasethree{} (\StwoEstrictthree) & \StwoEbasefour\\
S3 research runtimes & \nSthree & \SthreeEbaseone & \SthreeEbasetwo{} (\SthreeEstricttwo) & \SthreeEbasethree{} (\SthreeEstrictthree) & \SthreeEbasefour\\
\bottomrule
\end{tabular}
\end{table}

Practice and research separate already at RQ1 (Table~\ref{tab:e}). \nEthreePlus{} systems reach E3 or above, all but one of them in S1 or S2. \nEfour{} reach E4 (Hermes Agent, n8n, OpenClaw and OpenHands in S1; GitHub Copilot coding agent, OpenCode and Pi in S2), and none reaches E5. The strongest candidate, the Copilot coding agent, was lowered from E5 during verification because the multi-team scale and measurement window come only from vendor publications; independently checkable pull requests support E4 but no further. \SthreeEbaseone{} of the \nSthree{} research runtimes are at E1: they are evaluated on benchmarks, and we found no public report of anyone operating them. The exception, AgentScope, is a general agent framework rather than a recovery prototype, and its E3 rests on a deployment published by its own vendor; under the strict view it falls to E2. Under the strict view, E3+ falls from \nEthreePlus{} to \nEthreePlusStrict{} systems, but every E4 survives, because each rests on at least one operator-published source.

\subsection{RQ2: recovery mechanisms}
\begin{figure*}[t]
\centering
\includegraphics[width=.86\textwidth]{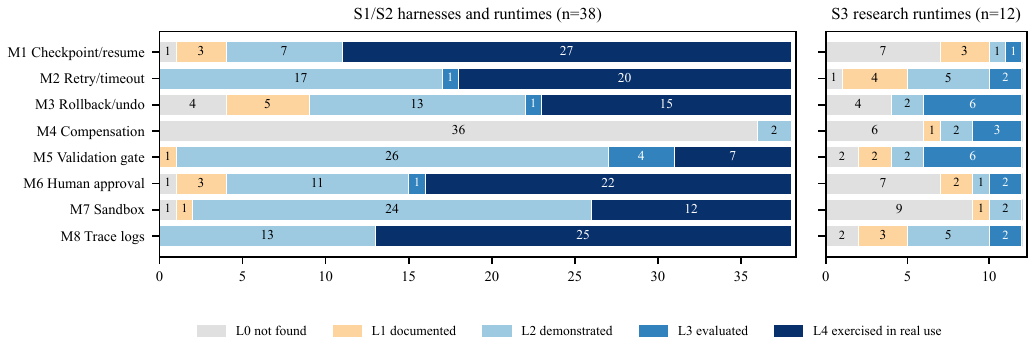}
\caption{Strongest public evidence per mechanism class: harnesses and runtimes (left, $n=38$) and research runtimes (right, $n=12$), same per-system scale. \textbf{Practice ships and exercises almost every class; research evaluates but never exercises. Compensation (M4) is the one class research covers and practice does not.}}
\label{fig:levels}
\end{figure*}

Figure~\ref{fig:levels} gives the distribution. Among the 38 S1/S2 systems, every class except compensation is at least documented for 34 or more systems, and most are demonstrated. Real-use (L4) reports are common for checkpoint/resume (\MoneLfour{} systems), trace logs (\MeightLfour), human approval (\MsixLfour) and retry/timeout (\MtwoLfour), less so for rollback (\MthreeLfour), sandboxing (\MsevenLfour) and validation gates (\MfiveLfour). \SoneanyLfour{} of \nSone{} S1 and \StwoanyLfour{} of \nStwo{} S2 systems have at least one L4 mechanism; L3 is rare in S1/S2, since harness projects seldom publish measurements of their own recovery.

\textbf{Exercised mechanisms rarely recover the run.} Of the \nLfour{} L4 reports, \nLfourrecovered{} describe the mechanism recovering the run, \nLfourpartial{} partial recovery, \nLfourfailed{} failure and \nLfourunclear{} are unclear (Fig.~\ref{fig:outcomes}). The pattern differs by mechanism. The three mechanisms that restore state fare worst: checkpoint/resume recovered in \MoneLfour{}~reports only \Monerec{} time (\Monefailed{} failed, \Monepartial{} partial), with typical reports of sessions that could be resumed only after deleting stale locks by hand or that resumed without their settings; rollback recovered once in \MthreeLfour{} reports, with failures that also discarded user work, as in the Aider and Kilo Code cases above. Trace logs fare best: \Meightrec{} of \MeightLfour{} reports recovered and none failed outright, because a persisted transcript or event log was what let the operator diagnose and repair the run. Human approval is split (\Msixrec{} recovered, \Msixfailed{} failed): interrupts that stopped a runaway loop sit beside approvals that were bypassed or, as with \emph{allow always} in Gemini CLI, authorised destruction.

This outcome distribution should not be read as a failure \emph{rate}: \nLfourIssueOnly{} of the \nLfour{} L4 labels rest on operator evidence that is only GitHub issues, and issues are filed when things break. What it does show is that the public record of exercised recovery is a record of incidents, and that the mechanisms most often relied on to restore state are the ones with the weakest record.

\subsection{RQ3: research guarantees versus practice}
\begin{table}[t]
\caption{Systems per mechanism at $\geq$L1, at L3 (research, $n=12$) and at L4 (practice, $n=38$). \textbf{Research evaluates rollback, validation and compensation; practice exercises resume, approval and logs.}}
\label{tab:rq3}
\centering\small
\begin{tabular}{lcccc}
\toprule
& \multicolumn{2}{c}{S3} & \multicolumn{2}{c}{S1/S2}\\
Mechanism & $\geq$L1 & L3 & $\geq$L1 & L4\\
\midrule
M1 checkpoint/resume & \MonegeOneSthree & \MoneLthreeSthree & \MonegeOneSOT & \MoneLfour\\
M2 retry/timeout & \MtwogeOneSthree & \MtwoLthreeSthree & \MtwogeOneSOT & \MtwoLfour\\
M3 rollback/undo & \MthreegeOneSthree & \MthreeLthreeSthree & \MthreegeOneSOT & \MthreeLfour\\
M4 compensation & \MfourgeOneSthree & \MfourLthreeSthree & \MfourgeOneSOT & \MfourLfour\\
M5 validation gate & \MfivegeOneSthree & \MfiveLthreeSthree & \MfivegeOneSOT & \MfiveLfour\\
M6 human approval & \MsixgeOneSthree & \MsixLthreeSthree & \MsixgeOneSOT & \MsixLfour\\
M7 sandbox & \MsevengeOneSthree & \MsevenLthreeSthree & \MsevengeOneSOT & \MsevenLfour\\
M8 trace logs & \MeightgeOneSthree & \MeightLthreeSthree & \MeightgeOneSOT & \MeightLfour\\
\bottomrule
\end{tabular}
\end{table}

Table~\ref{tab:rq3} sets the two populations side by side. Research runtimes concentrate their evaluated evidence on rollback, validation and compensation, and rarely address sandboxing or human approval, which nearly every harness ships (37 of 38 systems each) and operators exercise for \MsixLfour{} and \MsevenLfour{} systems respectively. Across the \nSthree{} S3 systems we recorded \nGuar{} stated recovery guarantees, \nGuarFormal{} of them with a formal definition or proof; none is supported by evidence above L3.

\textbf{Compensation is the clearest gap.} \MfourLthreeSthree{} research runtimes evaluate compensation of committed external effects (L3), and \MfourgeOneSthree{} at least document it. Among the 38 harnesses and runtimes, only \MfourgeOneSOT{} show anything at M4, and both (Langflow and Letta) are infrastructure clean-up, undoing the framework's own writes when a later step fails, not compensation of effects an agent's actions committed. No harness or runtime exposes semantic inverse actions or saga-style compensation. Harnesses handle side effects instead by preventing them (human approval, sandboxing) and by restoring local state (checkpoints, git-based undo), the very mechanisms whose exercised record in RQ2 is weakest.

\section{Discussion}
The two populations answer different questions. Research runtimes ask whether a recovery guarantee \emph{can} be provided and show that it can, on a benchmark. Harnesses ship pragmatic mechanisms and their users discover, in issues, where those mechanisms stop short. Neither side produces the evidence the other needs: no runtime reports a guarantee under real operation, and no harness measures its own recovery (L3 is nearly empty in S1/S2). This suggests three things. For researchers, the path to impact runs through exposure in an existing harness and an operator report, not through a further benchmark. For harness builders, resume and rollback are the mechanisms users rely on and report failing, so they merit the testing that validation gates and sandboxes already receive. For evaluators, the incident record in public issue trackers is a usable, if biased, source of field evidence about recovery, as shown here and by mining studies of agent repositories~\cite{robbes2026mining,zhang2026agenticbugs}.

\section{Threats to validity}
\textbf{What counts as an operator.} The frozen codebook treats any attributable individual, including a GitHub user filing an issue, as an operator. This is why the strict view barely moves the results (\nLfourStrict{} of \nLfour{} L4 labels survive; all E4 labels survive). Six of the seven E4 systems rest on a single operator's own report. A view restricted to named organisations would be stricter; we leave it to future work and treat single-operator E4 labels as the weakest part of the RQ1 result.

\textbf{Reporting bias.} Issue trackers over-represent failure. Outcome counts describe the public record, not failure rates, and successful recoveries are likely under-reported.

\textbf{Agent-performed coding and author review.} Both coding passes, adjudication and verification were performed by LLM-based agents; $\kappa$ values measure model--model agreement before author review, not human agreement. The two passes differed systematically, with Pass~B more often higher, and Pass~B searched less deeply (median 24 versus 34 evidence items; thin for Flowise and Semantic Raft). Mitigations were blinding, adjudication by codebook rule against archived snapshots, mandatory re-verification of every L4, E3+ and large disagreement, and an audit sample of agreed labels. The author subsequently checked every final label and outcome against the cited evidence and codebook. This review does not provide an independent human second coder or human inter-rater agreement.

\textbf{Construct and coverage.} The search was bounded; a missing label means \emph{not publicly found}. Two S3 systems (RoMa, Semantic Raft) were coded from abstracts because their full texts were inaccessible, so their levels are floors. One predeclared search step (issue search) was not run in Pass~B for Atomix. Two S3 runtimes have no released code.

\textbf{Conflict of interest.} The author co-authored one S3 runtime (RAC~\cite{perera2026rac}). It was coded from public sources only; a private repository was excluded, and its levels do not affect any conclusion.

\textbf{Time.} Agent harnesses change weekly. Labels hold at the cutoff and are tied to archived snapshots.

\section{Related work}
Studies of harness design show that the harness materially changes agent outcomes~\cite{yao2026harnessbench,pan2026nlah,zhang2026harnessdisclosure}, and anatomies of production harnesses describe their checkpoint, undo and sandbox subsystems from source code~\cite{barbaste2026harnessanatomy,liu2026diveclaudecode,rombaut2026scaffold}; we add evidence of how those subsystems behave in use. The AI Agent Index documents deployed agents from public information~\cite{staufer2026agentindex}, and AgentStripes grades the evidence behind 40 maturity capabilities of public agent repositories~\cite{naik2026agentstripes}; our ladder is recovery-specific and spans research runtimes. Mastouri et al.\ measure, for persistence in LangGraph deployments, the kind of guarantee-versus-practice gap we map across eight mechanisms~\cite{mastouri2026builttoremember}, and Wu et al.\ show that correct rollback is not secure recovery~\cite{wu2026safetoresume}. Fault-injection benchmarks probe agent recovery under controlled faults~\cite{jia2026masfire,gupta2026reliabilitybench,zhang2026agentchaosbench}; failure taxonomies classify how agents fail~\cite{cemri2025mast,shah2026faults}. Practitioner studies name reliability as the leading production concern~\cite{pan2025map}.

\section{Conclusion}
We graded the public evidence for eight recovery mechanisms across 50 agent harnesses, runtimes and research systems. Recovery is exercised almost everywhere in practice, yet only \nLfourrecovered{} of \nLfour{} real-use reports describe a full recovery, and the guarantees research evaluates, compensation above all, are not shown in any harness. Agent recovery should be judged by what happens when operators use it, so research guarantees need to reach a harness, and harness mechanisms need measuring before users find their limits. Replication package: \url{https://doi.org/10.5281/zenodo.24122560}.

\bibliographystyle{IEEEtran}
\bibliography{references}
\end{document}